# Uncertain Time-Series Similarity: Return to the Basics

Michele Dallachiesa, Besmira Nushi, Katsiaryna Mirylenka, and Themis Palpanas

University of Trento
{dallachiesa,kmirylenka,themis}@disi.unitn.eu, besmira.nushi@studenti.unitn.it

## ABSTRACT

In the last years there has been a considerable increase in the availability of continuous sensor measurements in a wide range of application domains, such as Location-Based Services (LBS), medical monitoring systems, manufacturing plants and engineering facilities to ensure efficiency, product quality and safety, hydrologic and geologic observing systems, pollution management, and others.

Due to the inherent imprecision of sensor observations, many investigations have recently turned into querying, mining and storing uncertain data. Uncertainty can also be due to data aggregation, privacy-preserving transforms, and error-prone mining algorithms.

In this study, we survey the techniques that have been proposed specifically for modeling and processing uncertain time series, an important model for temporal data. We provide an analytical evaluation of the alternatives that have been proposed in the literature, highlighting the advantages and disadvantages of each approach, and further compare these alternatives with two additional techniques that were carefully studied before. We conduct an extensive experimental evaluation with 17 real datasets, and discuss some surprising results, which suggest that a fruitful research direction is to take into account the temporal correlations in the time series. Based on our evaluations, we also provide guidelines useful for the practitioners in the field.

## 1. INTRODUCTION

In the last decade there has been a dramatic explosion in the availability of measurements in a wide range of application domains, including traffic flow management, meteorology, astronomy, remote sensing, and object tracking. Applications in the above domains usually organize these sequential measurements into time series, i.e., sequences of data points ordered along the temporal dimension, making time series a data type of particular importance.

Several studies have recently focused on the problems of processing and mining time series with incomplete, imprecise and even misleading measurements [7, 14, 25, 26, 28]. Uncertainty in time series may occur for different reasons, such as the inherent imprecision of sensor observations, or privacy-preserving transformations. The following two examples illustrate these two cases:

- Personal information contributed by individuals and corporations is steadily increasing, and there is a parallel growing interest in applications that can be developed by mining these datasets, such as location-based services and social network applications. In these applications privacy is a major concern, addressed by various privacy-preserving transforms [2, 11, 20], which introduce data uncertainty. The data can still be mined and queried, but it requires a re-design of the existing methods in order to address this uncertainty.

- In manufacturing plants and engineering facilities, sensor networks are being deployed to ensure efficiency, product quality and safety [14]: unexpected vibration patterns in production machines, or changes in chemical composition in industrial processes, are used to predict failures, suggesting repairs or replacements. The same is true in environmental science [12], where sensor networks are used in hydrologic and geologic observing systems, pollution management in urban settings, and application of water and fertilizers in precision agriculture. In transportation, sensor networks are employed to monitor weather and traffic conditions, and increase driving safety [21]. However, sensor readings are inherently imprecise because of the noise introduced by the equipment itself [7]. This translates to time series with uncertain values, and addressing this uncertainty can provide better results in terms of quality and efficiency.

While the problem of managing and processing uncertain data has been studied in the traditional database literature since the 80's [3], the attention of researchers was only recently focused on the specific case of uncertain time series. Two main approaches have emerged for modeling uncertain time series. In the first, a probability density function (pdf) over the uncertain values is estimated by using some a priori knowledge [30, 29, 23]. In the second, the uncertain data distribution is summarized by repeated measurements (i.e., samples) [5].

In this study, we revisit the techniques that have been proposed under these two approaches, with the aim of determining their advantages and disadvantages. This is the first study to undertake a rigorous comparative evaluation of the techniques proposed in the literature for similarity matching of uncertain time series. The importance of such a study is underlined by two facts: first, the widespread existence of uncertain time series; and second, the observation that similarity matching serves as the basis for developing various more complex analysis and mining algorithms. Therefore, acquiring a deep understanding of the techniques proposed in this area is essential for the further development of the field of uncertain time series processing, and the applications that are built on top of it [27, 15, 17].

Our evaluation reveals the effectiveness of the techniques that have been proposed in the literature under different scenarios. In the experiments, we stress-test the different techniques both in situations for which they were designed, as well as in situations that fall outside their normal operation (e.g., unknown distributions of the uncertain values). In the latter case, we wish to establish how strong the assumptions behind the design principles of each technique are, and to what extent these techniques can produce reliable and stable results, when these assumptions no longer hold. We note that such situations do arise in practice, where it is not always possible to know the exact data characteristics of the uncertain time series.

Furthermore, we describe additional similarity measures for uncertain time series, inspired by the moving average, namely Uncertain Moving Average (UMA), and Uncertain Exponential Moving Average (UEMA). Even though these similarity measures are very simple, previous studies had not considered them. However, the experimental evaluation shows that they perform better than the more sophisticated techniques that have been proposed in the literature. The reason lies in the fact that UMA and UEMA incorporate some of the information inherent in the *sequence* of points in the time series, thus, taking a step back from the independence assumption of the other techniques. We argue that these measures, which are simple and computationally efficient, should serve as the baseline for the problem of similarity matching in uncertain time series.

Moreover, we make sure that the results of our experiments are completely reproducible. Therefore, we make publicly available the source code for all the algorithms used in our experiments, as well as the datasets upon which we tested them[1].

In summary, we make the following contributions.

- We review the state of the art techniques for similarity matching in uncertain time series, and analytically evaluate them. Our analysis serves as a single-stop comparison of the proposed techniques in terms of requirements, input data assumptions, and applicability to different situations.

- We propose a methodology for comparing these techniques, based on the similarity matching task. This methodology provides a common ground for the fair comparison of all the techniques.

- We perform an extensive experimental evaluation, using 17 real datasets from diverse domains. In our experiments, we evaluate the techniques using a multitude of different conditions, and input data characteristics. Moreover, we stress-test the techniques by evaluating their performance on datasets for which they have not been designed to operate.

- We describe and evaluate two additional similarity measures for uncertain time series, that were not studied in this context before. These measures are based on moving average, and one of them also employs exponential decaying. We demonstrate that the new measures achieve better performance than the similarity measures in the literature, which is an unexpected result.

- Finally, we provide a discussion of the results, and complement this discussion with thoughts on interesting research directions, as well as useful guidelines for the practitioners in the field.

The rest of this paper is structured as follows. In Section 2 we survey the principal representations and distance measures proposed for similarity matching of uncertain time series. In Section 3, we analytically compare the methods proposed for uncertain time series modeling, and in Section 4, we present the experimental comparison. We describe new measures of similarity matching in Section 5, and evaluate their performance in relation to the other measures. Finally, in Section 6 we summarize the results, and Section 7 concludes the study.

## 2. SIMILARITY MATCHING FOR UNCERTAIN TIME SERIES

Time series are sequences of points, typically real valued numbers, ordered along the temporal dimension. We assume constant sampling rates and discrete timestamps. Formally, a time series $S$ is defined as $S = <s_1, s_2, ..., s_n>$ where $n$ is the length of $S$, and $s_i$ is the real valued number of $S$ at timestamp $i$. Where not specified otherwise, we assume normalized time series with zero mean and unit variance. Notice that normalization is a preprocessing step that requires particular care to address specific situations [16].

In this study, we focus on uncertain time series where uncertainty is localized and limited to the points. Formally, an uncertain time series $T$ is defined as a sequence of random variables $<t_1, t_2, ..., t_n>$ where $t_i$ is the random variable modeling the real valued number at timestamp $i$. All the three models we review and compare fit under this general definition.

The problem of similarity matching has been extensively studied in the past [4, 10, 22, 13, 8, 19, 18, 16] : given a user-supplied query sequence, a similarity search returns the most similar time series according to some distance function. More formally, given a collection of time series $C = \{S_1, ..., S_N\}$, where $N$ is the number of time series, we are interested in evaluation the range query function $RQ(Q, C, \epsilon)$:

$$RQ(Q, C, \epsilon) = \{S | S \in C | distance(Q, S) \leq \epsilon\} \quad (1)$$

In the above equation, $\epsilon$ is a user-supplied distance threshold. A survey of representation and distance measures for time series can be found in [9].

A similar problem arises also in the case of uncertain time series, and the problem of probabilistic similarity matching has been introduced in the last years. Formally, given a collection of uncertain time series $C = \{T_1, ..., T_N\}$, we are

---

[1]Source code and datasets:
http://disi.unitn.eu/~dallachiesa/codes/uncertts/

interested in evaluation the probabilistic range query function $PRQ(Q, C, \epsilon, \tau)$:

$$PRQ(Q, C, \epsilon, \tau) = \{T | T \in C | Pr(distance(Q, S) \leq \epsilon) \geq \tau\} \quad (2)$$

In the above equation, $\epsilon$ and $\tau$ are the user-supplied distance threshold and the probabilistic threshold, respectively.

In the recent years three techniques have been proposed to evaluate $PRQ$ queries, namely MUNICH[2] [5], PROUD [29], and DUST [23]. As we discuss below, these methods assume that neighboring points of the time series are independent, i.e., the point at timestamp $i$ is independent from the point at timestamp $i+1$. Evidently, this is a simplifying assumption, since in real-world datasets neighboring points are correlated. We revisit this issue in the following sections.

We now discuss each one of the above three techniques in more detail.

## 2.1 MUNICH

In [5], uncertainty is modeled by means of repeated observations at each timestamp, as depicted in Figure 2.

Assuming two uncertain time series, $X$ and $Y$, MUNICH proceeds as follows. First, the two uncertain sequences $X, Y$ are materialized to all possible certain sequences: $TS_X = \{<v_{11}, ..., v_{n1}>, ..., <v_{1s}, ..., v_{ns}>\}$ (where $v_{ij}$ is the $j$-th observation in timestamp $i$), and similarly for $Y$ with $TS_Y$. Thus, we have now defined $TS_X, TS_Y$. The set of all possible distances between $X$ and $Y$ is then defined as follows:

$$dists(X, Y) = \{L^p(x, y) | x \in TS_X, y \in TS_Y\} \quad (3)$$

The uncertain $L^p$ distance is formulated by means of counting the feasible distances:

$$Pr(distance(X, Y) \leq \epsilon) = \frac{|\{d \in dists(X, Y) | d \leq \epsilon\}|}{|dists(X, Y)|} \quad (4)$$

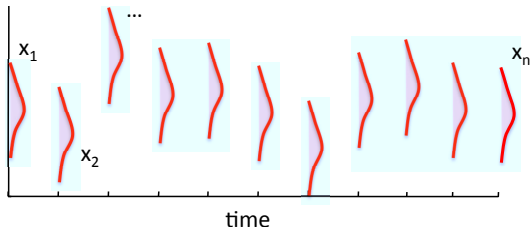

Figure 1: Example of uncertain time series $X = \{x_1, ..., x_n\}$ modeled by means of $pdf$ estimation.

Once we compute this probability, we can determine the result set of PRQs similarity queries by filtering all uncertain sequences using Equation 4.

Note that the naive computation of the result set is infeasible, because of the very large space that leads to an exponential computational cost: $|dists(X, Y)| = s_X^n s_X^n$, where $s_X, s_Y$ are the number of samples at each timestamp of $X, Y$, respectively, and $n$ is the length of the sequences.

[2]We will refer to this method as MUNICH (it was not explicitly named in the original paper), since all the authors were affiliated with the University of Munich.

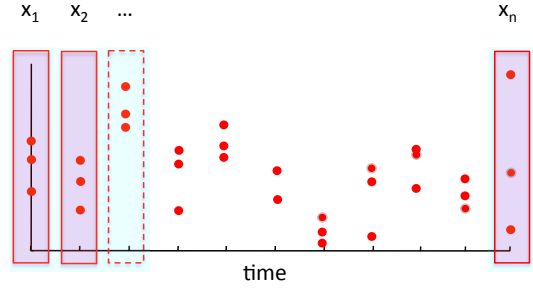

Figure 2: Example of uncertain time series $X = \{x_1, ..., x_n\}$ modeled by means of repeated observations.

Efficiency can be ensured by upper and lower bounding the distances, and summarizing the repeated samples using minimal bounding intervals [5]. This framework has been applied to Euclidean and Dynamic Time Warping (DTW) [6] distances and guarantees no false dismissals in the original space [5].

## 2.2 PROUD

In [29], an approach for processing queries over PRObabilistic Uncertain Data streams (PROUD) is presented. Inspired by the Euclidean distance, the PROUD distance is modeled as the sum of the differences of the streaming time series random variables, where each random variable represents the uncertainty of the value in the corresponding timestamp. This model is illustrated in Figure 1.

Given two uncertain time series $X, Y$, their distance is defined as:

$$distance(X, Y) = \sum_i D_i^2 \quad (5)$$

where $D_i = (x_i - y_i)$ are random variables, as shown in Figure 3.

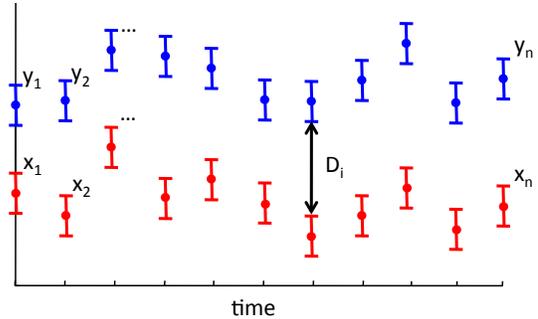

Figure 3: The probabilistic distance model.

According to the central limit theorem, we have that the cumulative distribution of the distances approaches a normal distribution:

$$distance(X, Y)_{norm} = \frac{distance(X, Y) - \sum_i E[d_i^2]}{\sqrt{\sum_i Var[D_i^2]}} \quad (6)$$

The normalized distance follows a standard normal distribution, thus we can obtain the normal distribution of the original distance as follows:

$$distance(X,Y) \propto N(\sum_i E[D_i^2], \sum_i Var[D_i^2]) \quad (7)$$

The interesting result here is that, regardless of the data distribution of the random variables composing the uncertain time series, the cumulative distribution of their distances (1) is defined similarly to their Euclidean distance and (2) approaches a normal distribution. Recall that we want to answer PRQs similarity queries. First, given a probability threshold $\tau$ and the cumulative distribution function (*cdf*) of the normal distribution, we compute $\epsilon_{limit}$ such that:

$$Pr(distance(X,Y)_{norm} \leq \epsilon_{limit}) \geq \tau \quad (8)$$

The *cdf* of the normal distribution can be formulated in terms of the well-known *error-function*, and $\epsilon_{limit}$ can be determined by looking up the statistics tables. Once we have $\epsilon_{limit}$, we proceed by computing also the normalized $\epsilon$:

$$\epsilon_{norm}(X,Y) = \frac{\epsilon^2 - E[distance(X,Y)]}{\sqrt{Var[distance(X,Y)]}} \quad (9)$$

Then, we have that if a candidate uncertain series $Y$ satisfies the inequality:

$$\epsilon_{norm}(X,Y) \geq \epsilon_{limit} \quad (10)$$

then the following equation holds:

$$Pr(distance(X,Y)_{norm} \leq \epsilon_{norm}(X,Y)) \geq \tau \quad (11)$$

Therefore, $Y$ can be added to the result set. Otherwise, it is pruned away. This distance formulation is statistically sound and only requires knowledge of the general characteristics of the data distribution, namely, its mean and variance.

## 2.3 DUST

In [23], the authors propose a new distance measure, DUST, that compared to MUNICH, does not depend on the existence of multiple observations and is computationally more efficient. Similarly to [29], DUST is inspired by the Euclidean distance, but works under the assumption that all the time series values follow some specific distribution.

Given two uncertain time series $X, Y$, the distance between two uncertain values $x_i, y_i$ is defined as the distance between their true (unknown) values $r(x_i), r(y_i)$: $dist(x_i, y_i) = L^1(r(x_i), r(y_i))$. This distance can then be used to define a function $\phi$ that measures the similarity of two uncertain values:

$$\phi(|x_i - y_i|) = Pr(dist(0|r(x_i) - r(y_i)|) = 0) \quad (12)$$

This basic similarity function is then used inside the *dust* dissimilarity function:

$$\begin{aligned} dust(x,y) &= \sqrt{-\log(\phi(|x-y|)) - k} \\ &\text{with} \\ k &= -\log(\phi(0)) \end{aligned}$$

The constant $k$ has been introduced to support reflexivity. Once we have defined the *dust* distance between uncertain values, we are ready to extend it to the entire sequences:

$$DUST(X,Y) = \sqrt{\sum_i dust(x_i, y_i)^2} \quad (13)$$

The handling of uncertainty has been isolated inside the $\phi$ function, and its evaluation requires to know exactly the data distribution. In contrast to the techniques we reviewed earlier, the DUST distance is a real number that measures the dissimilarity between uncertain time series. Thus, it can be used in all mining techniques for certain time series, by simply substituting the existing distance function.

Finally, we note that DUST is equivalent to the Euclidean distance, in the case where the error of the time series values follows the normal distribution.

## 3. ANALYTICAL COMPARISON

In this section, we compare the three models of similarity matching for uncertain time series, namely, MUNICH, PROUD and DUST, along the following dimensions: uncertainty models used and assumptions made by the algorithms; type of distance measures; and type of similarity queries.

### 3.1 Uncertainty Models and Assumptions

All three techniques we have reviewed are based on the assumption that the values of the time series are independent from one another. That is, the value at each timestamp is assumed to be independently drawn from a given distribution. Evidently, this is a simplifying assumption, since neighboring values in time series usually have a strong temporal correlation.

The main difference between MUNICH and the other two techniques is that MUNICH represents the uncertainty of the time series values by recording multiple observations for each timestamp. This can be thought of as sampling from the distribution of the value errors. In contrast, PROUD and DUST consider each value of time series to be a continuous random variable following a certain probability distribution.

The amount of preliminary information, i.e. a priori knowledge of the characteristics of the time series values and their errors, varies greatly among the techniques. MUNICH does not need to know the distribution of the time series values, or the distribution of the value errors. It simply operates on the observations available at each timestamp.

On the other hand, PROUD and DUST need to know the distribution of the error at each value of the data stream. In particular, PROUD requires to know the standard deviation of the uncertainty error, and a single observed value for each timestamp. PROUD assumes that the standard deviation of the uncertainty error remains constant across all timestamps.

DUST uses the largest amount of information among the three techniques. It takes as input a single observed value of the time series for each timestamp, just like PROUD. In addition, DUST needs to know the distribution of the uncertainty error at each time stamp, as well as the distribution of the values of the time series. This means that, in contrast to PROUD, DUST can take into account mixed distributions for the uncertainty errors (albeit, they have to be explicitly provided in the input).

Overall, we observe that the three techniques make different initial assumptions about the amount of information available for the uncertain time series, and have different input requirements. Consequently, when deciding which technique to use, users should take into account the information available on the uncertainty of the time series to be processed.

## 3.2 Type of Distance Measures

All the considered techniques use some variation of the Euclidean distance. MUNICH and PROUD use this distance in a pretty straightforward manner. Moreover, MUNICH and DUST can be employed to compute the Dynamic Time Warping distance [24], which is a more flexible distance measure.

DUST is a new type of distance, specifically designed for uncertain time series. In other words, DUST is not a similarity matching technique per se, but rather a new distance measure. It has been shown that DUST is proportional to the Euclidean distance in the cases where the value errors are normally distributed [23]. Moreover, the authors of [23] note that if all the value errors follow the same distribution, then it is better to use the Euclidean distance. DUST becomes useful when the value errors are modeled by multiple error distributions.

## 3.3 Type of Similarity Queries

MUNICH and PROUD are designed for answering probabilistic range queries (defined in Section 2). DUST being a distance measure, it can be used to answer top-k nearest neighbor queries, or perform top-k motif search.

MUNICH and PROUD solve the similarity matching problem that is described by Equation 8, resulting to a set of time series that belong to the answer with a certain probability, $\tau$. On the other hand, DUST produces a single value that is an exact (i.e., not probabilistic) distance between two uncertain time series.

In Section 4, we describe the methodology we used in order to compare all three techniques using the same task, that of similarity matching.

## 4. COMPARATIVE STUDY

In this section, we present the experimental evaluation of the three techniques. We first describe the methodology and datasets used, and then discuss the results of the experiments.

All techniques were implemented in C++, and the experiments were run on a PC with a 2.13GHz CPU and 4GB of RAM.

The source code for all the algorithms used in our experiments, as well as the datasets upon which we tested them are publicly available[1].

## 4.1 Experimental Setup

### 4.1.1 Datasets

Similarly to [5, 29, 23], we used existing time series datasets with exact values as the ground truth, and subsequently introduced uncertainty through perturbation. Perturbation models errors in measurements, and in our experiments we consider *uniform*, *normal* and *exponential* error distributions with zero mean and varying standard deviation within interval [0.2, 2.0].

We considered 17 real datasets from the UCR classification datasets collection [1], representing a wide range of application domains: 50words, Adiac, Beef, CBF, Coffee, ECG200, FISH, FaceAll, FaceFour, Gun_Point, Lighting2, Lighting7, OSULeaf, OliveOil, SwedishLeaf, Trace, and synthetic_control. The training and testing sets were joined together, and we obtained on average 502 time series of length 290 per dataset. We stress the fact that each dataset contains several time series instances.

Since DUST requires to know the distribution of values of the time series, and additionally makes the assumption that this distribution is uniform [23], we tested the datasets to check if this assumption holds. According to the Chi-square test, the hypothesis that the datasets follow the uniform distribution was rejected (for all datasets) with confidence level $\alpha = 0.01$. Evidently, the above assumption does not hold on all datasets, however DUST still needs it in order to operate.

### 4.1.2 Comparison Methodology

In our evaluation, we consider all three techniques, namely, MUNICH, PROUD, and DUST, and we additionally compare to Euclidean distance. When using Euclidean distance, we do not take into account the distributions of the values and their errors: we just use a single value for every timestamp, and compute the traditional Euclidean distance based on these values.

The goal of our evaluation is to compare the performance of the different techniques on the same task. Observe that we cannot use the top-k search task for this comparison. The reason is that the MUNICH and PROUD techniques have a notion of probability (Equation 2). This means that these techniques can produce different rankings when the threshold $\varepsilon$ changes. For example, assume that we increase $\varepsilon$ (maintaining $\tau$ fixed). Then the ordering of the time series in a top-k ranking may change, since the probability that the time series are similar within distance $\varepsilon_1 \geq \varepsilon$ may increase. Thus, in the case of uncertain time series, MUNICH and PROUD might produce very different top-k answers even if $\varepsilon$ varies a little. This, in turn, means that the top-k task is not suitable for comparing the three techniques.

We instead perform the comparison using the task of time series similarity matching. Even though DUST is not a similarity matching technique (like PROUD and MUNICH), it can still be used to find similar time series, when we specify a maximum threshold on the distance between time series. In [23], the evaluation of DUST was based on top-k similar time series. However, we note that this problem includes the problem of similarity matching [9], where the most similar time series form the answer to the top-k query.

Following the above discussion, in order to perform a fair comparison we need to specify distance thresholds for all three techniques. This translates to finding equivalent thresholds $\varepsilon$ for each one of the techniques. We proceed as follows.

Since the distances in MUNICH and PROUD are based on the Euclidean distance, we will use the same threshold for both methods, $\varepsilon_{eucl}$. Then, we calculate an equivalent threshold for DUST, $\varepsilon_{dust}$. Given a query $q$ and a dataset $C$, we identify the 10th nearest neighbor of $q$ in $C$. Let that be time series $c$. We define $\varepsilon_{eucl}$ as the Euclidean distance on the observations between $q$ and $c$ and $\varepsilon_{dust}$ as the DUST distance between $q$ and $c$. This procedure is repeated for every query $q$.

The quality of results of the different techniques is evaluated by comparing the query results to the ground truth. We performed experiments for each dataset separately, using each one of the time series as a query and performing a similarity search. In the graphs, we report the averages of all these results, as well as the 95% confidence intervals[3].

## 4.2 Quality Performance

In order to evaluate the quality of the results, we used the two standard measures of *recall* and *precision*. Recall is defined as the percentage of the truly similar uncertain time series that are found by the algorithm. Precision is the percentage of similar uncertain time series identified by the algorithm, which are truly similar. Accuracy is measured in terms of $F_1$ score to facilitate the comparison. The $F_1$ score is defined by combining precision and recall:

$$F_1 = 2 * \frac{precision * recall}{precision + recall} \qquad (14)$$

We verify the results with the exact answer using the ground truth, and compare the results with the algorithm output (as described in Section 4.1.2).

### 4.2.1 Accuracy

The first experiment represents a special case with restricted settings. This was necessary to do, because the computational cost of MUNICH was prohibitive for a full scale experiment. We compare MUNICH, PROUD, DUST and Euclidean on the Gun_Point dataset, truncating it to 60 time series of length 6. For each timestamp, we have 5 samples as input for MUNICH. Results are averaged on 5 random queries. For both MUNICH and PROUD we are using the optimal probabilistic threshold, $\tau$, determined after repeated experiments. Distance thresholds are chosen (according to Section 4.1.2) such that in the ground truth set they return exactly 10 time series.

The results with Gaussian error (refer to Figure 4(a)) show that all techniques perform well ($F_1 > 80\%$) when the standard deviation of the errors is low ($\sigma = 0.2$), with MUNICH being the best performer ($F_1 = 88\%$). However, as the standard deviation increases to 2, the accuracy of all techniques decreases. This is expected, since a larger standard deviation means that the time series have more uncertainty. The behavior of MUNICH though, is interesting: its accuracy falls sharply for $\sigma > 0.6$.

This trend was verified also with uniform and exponential error distributions, as reported in Figures 4(b) and 4(c). With exponential error, the performance of MUNICH is slightly better than with normal, or uniform error distributions. However, MUNICH still performs much worse than PROUD and DUST for $\sigma > 0.6$.

Figure 5(a) shows the results of the same experiment, but just for PROUD, DUST, and Euclidean. In this case (and for all the following experiments), we report the average results over the full time series for all datasets. Once again, the error distribution is normal, and PROUD is using the optimal threshold, $\tau$, for every value of the standard deviation.

The results show that there is virtually no difference among the different techniques. This observation holds across the entire range of standard deviations that we tried ($0.2 \leq \sigma \leq 2$).

The results for the uniform and exponential distributions are very similar, and reported in Figures 5(b) and 5(c). With both uniform and exponential errors, PROUD performs slightly better for $\sigma = 0.2$, and its performance drops slightly below DUST and Euclidean for larger error standard deviations.

With uniform error, the accuracy of DUST drops by nearly 10% for $\sigma = 0.2$ (refer to Figures 5(b)). This apparently insignificant observation turned out to be due to how the DUST lookup tables are determined: When the error is uniformly distributed, $\phi(|x_i - y_i|)$ may be equal to zero in some cases. Consequently, $dust(x,y)$ cannot be evaluated for these cases, as it degenerates to the logarithm of zero. We tried to solve this technical problem by adding two tails to the uniform error, so that the error probability density function is never exactly zero. This workaround proved useful, but did not completely solve the problem.

### 4.2.2 Precision and Recall

In order to better understand the behavior of the different techniques, we take a closer look at precision and recall. Figures 6(a) and 6(b) show the precision and recall for PROUD, as a function of the error standard deviation, when the distribution of the error follows a uniform, a normal, and an exponential distribution. PROUD is using the optimal threshold, $\tau$, for *every* value of the standard deviation.

The graphs show that recall always remains relatively high (between $63\% - 83\%$). On the contrary, precision is heavily affected, decreasing from 70% to a mere 16% as standard deviation increases from 0.2 to 2. Therefore, processing uncertain time series with an increasing standard deviation in their error does not have a significant impact on the false positives. However, this leads to many false negatives, which may be an undesirable effect.

The corresponding results for DUST are shown in Figures 7(a) and 7(b). We observe the same trends as before, the only difference being that DUST achieves slightly better precision, but lower recall.

### 4.2.3 Mixed Error Distributions

While in all previous experiments the error distribution is constant across all the values of a time series, in this experiment we evaluate the accuracy of PROUD, DUST, and Euclidean when we have different error distributions present in the same time series (Figure 8). Each time series has been perturbed with normal error, but of varying standard deviation. Namely, the error for 20% of the values has standard deviation 1, and the rest 80% has standard deviation 0.4.

We note that this is a case that PROUD cannot handle, since it does not have the ability to model different error distributions within the same time series (in this experiment, PROUD was using a standard deviation setting of 0.7). Therefore, PROUD does not produce better results than Euclidean. On the other hand, DUST is taking into account these variations of the error, and achieves a slightly improved accuracy (3% more than PROUD and Euclidean).

We also conducted the same experiment by changing the following settings: (i) inform DUST (wrongly) that the standard deviation is 0.7, and (ii) perturb a time series with a mixture of uniform, normal, and exponential distributions (this situation cannot be handled by PROUD).

---
[3]Please note that the results we report are not directly comparable to those in the original papers. In our study, we use a different experimental setup, in order to make possible the comparison of the three techniques.

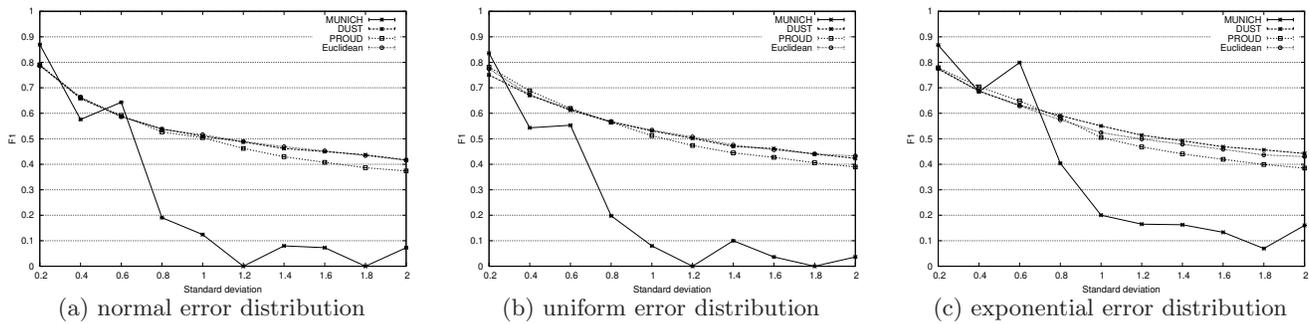

**Figure 4:** $F_1$ score for MUNICH, PROUD, DUST and Euclidean on Gun_Point truncated dataset, when varying the error standard deviation: normal error distribution (left), uniform (center), exponential (right).

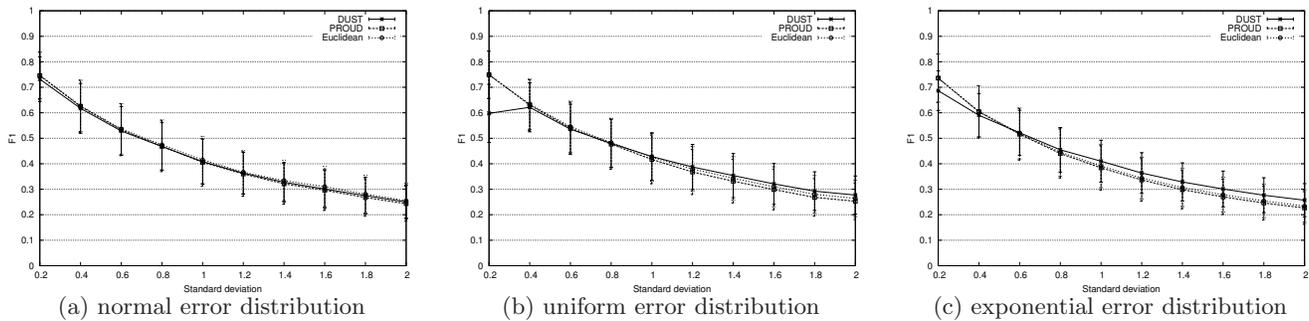

**Figure 5:** $F_1$ score for PROUD, DUST and Euclidean, averaged over all datasets, when varying the error standard deviation: normal error distribution (left), uniform (center), exponential (right).

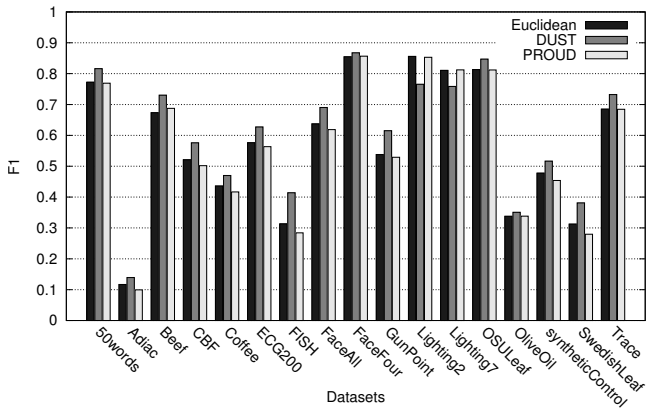

**Figure 8:** $F_1$ score for PROUD, DUST, and Euclidean on all the datasets with mixed error distribution (normal), 20% with standard deviation 1.0, and 80% with standard deviation 0.4.

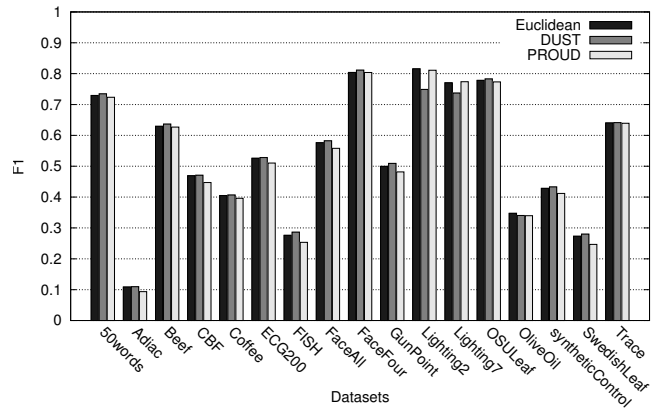

**Figure 9:** $F_1$ score for PROUD, DUST, and Euclidean on all the datasets with mixed error distribution (uniform, normal, and exponential), 20% with standard deviation 1.0, and 80% with standard deviation 0.4.

As shown in Figures 9 and 10, in both these experiments the accuracy of all techniques (PROUD, DUST, and Euclidean) is almost the same, and consistently lower for the second experiment. These results indicate that in situations where we do not have enough, or accurate information on the distribution of the error, PROUD and DUST do not offer an advantage when compared to Euclidean.

### 4.3 Time Performance

In Figure 11, we report the CPU time per query for the normal error distribution when varying the error standard deviation in the range $[0.2, 2.0]$. The results for uniform and exponential distributions are very similar, and omitted for brevity.

The graph shows that the standard deviation of the normal distribution only slightly affects performance for DUST. As expected, the execution time for Euclidean is not affected at all when the standard deviation for the error of the uncertain time series varies, and exhibits the best time performance of all techniques.

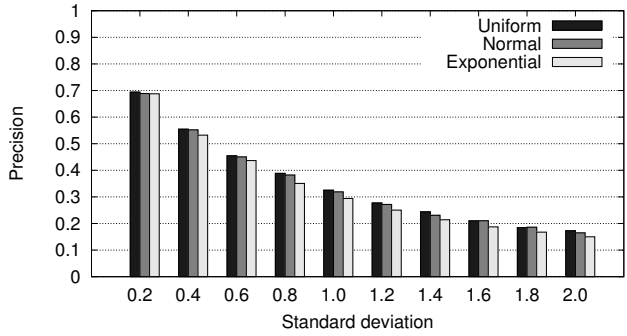
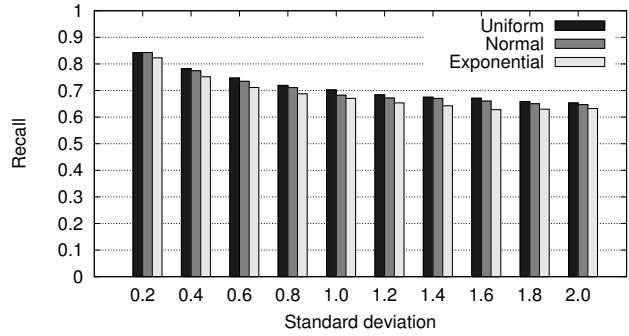

(a)  (b)

**Figure 6:** Precision and recall for PROUD, averaged over all datasets, when varying error standard deviation and error distribution.

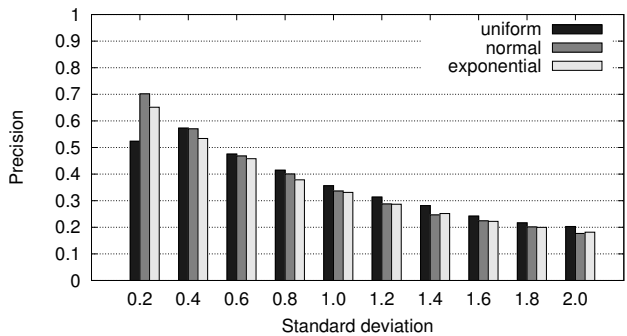
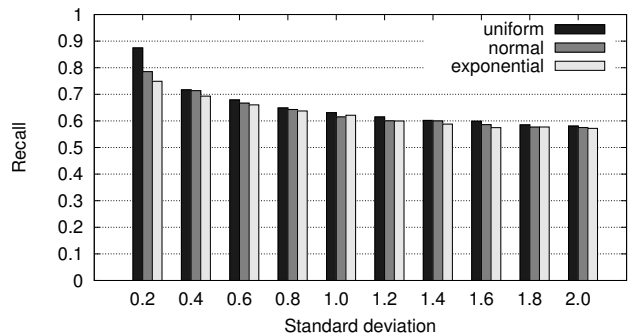

(a)  (b)

**Figure 7:** Precision and recall for DUST, averaged over all datasets, when varying error standard deviation and error distribution.

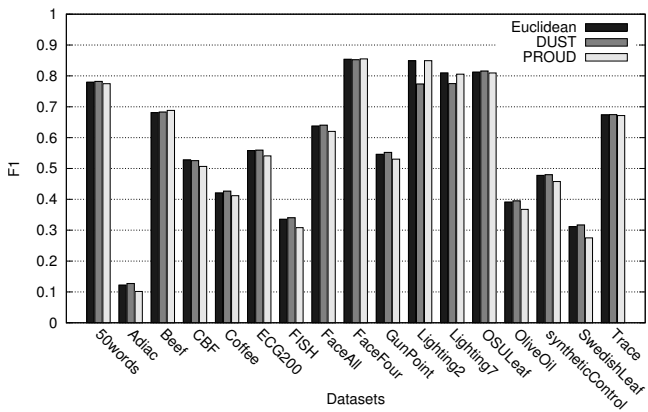

**Figure 10:** $F_1$ score for PROUD, DUST, and Euclidean on all the datasets with mixed error distribution: normal, with standard deviation erroneously reported as constant $0.7$.

We note that for PROUD we did not use the wavelet synopsis, since we did not use any summarization technique for the other techniques either. However, it is possible to apply PROUD on top of a Haar wavelet synopsis. This results in CPU time for PROUD that is equal or less to the CPU time of Euclidean, while maintaining high accuracy [29].

We did not include the time performance for MUNICH in this graph, because it is orders of magnitude more expensive than the other techniques (i.e., in the order of minutes).

In Figure 12, we report the CPU time per query for the normal error distribution when varying the time series length between 50 and 1000 time points. Time series of different lengths have been obtained resampling the raw sequences. The graph shows that the time grows linearly to the time series length. The results for uniform and exponential distributions are very similar, and omitted for brevity.

## 5. MOVING AVERAGE FOR UNCERTAIN TIME SERIES

The moving average is among the simplest filters for noise reduction in signal processing. In this section, we compare some basic adaptations of the moving average to the DUST and Euclidean distances, and evaluate their effectiveness. We note that similar to the Euclidean and DUST distances, it does not provide any quality guarantees in the context of uncertain time series similarity matching. (In contrast, MU-NICH and PROUD provide an additional probabilistic measure of certainty for the computed similarity value.) Nevertheless, the measures we describe below take a first step away from the assumption, which the techniques we examined so far make, that neighboring points in a time series are independent.

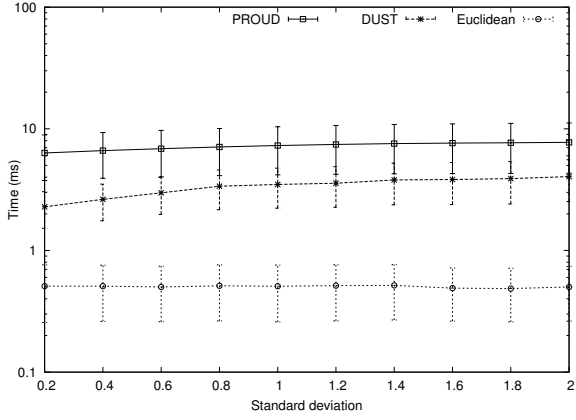

Figure 11: Average time per query for PROUD, DUST, and Euclidean, averaged over all datasets, when varying the error standard deviation with normal error distribution.

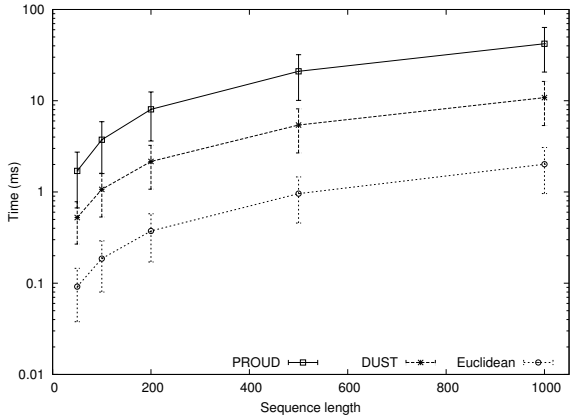

Figure 12: Average time per query for PROUD, DUST, and Euclidean, averaged over all datasets, when varying the time series length with normal error distribution.

### 5.1 Neighborhood-Aware Models

Given a series of noisy measurements $S = <v_1, v_2, ..., v_n>$, the moving average of these measurements $S_m$ is obtained by substituting each value $v_i$ with $m_i$, defined as the average of values $v_{i-w}, ..., v_i, v_{i+w}$:

$$m_i = \frac{\sum_{j=i-w}^{i+w} v_j}{2w+1} \quad (15)$$

where $w$ is a user-defined parameter that defines the window width $2w+1$ to be considered in the average. In the moving average, all samples are weighted equally.

A variant of the moving average, namely the exponential moving average, has been introduced to weigh more the nearest neighbors of the current value, through an exponential decaying factor. The exponential moving average of sequence $S$, $S_e$, is obtained by substituting each value $v_i$ with $e_i$, defined as follows:

$$e_i = \frac{\sum_{j=i-w}^{i+w} v_j e^{-\lambda|j-i|}}{\sum_{j=i-w}^{i+w} e^{-\lambda|j-i|}} \quad (16)$$

where $\lambda$ controls the exponential decaying factor.

The above two moving average filters require no a priori knowledge of the data distribution, and their parameters are intuitive and easy to tune, thus making these techniques widely adopted in the real world. In the next paragraphs, we introduce two variants of the moving and exponential moving averages that exploit the a priori knowledge of the error standard deviation.

Intuitively, we can weigh less the observations drawn from random variables that exhibit larger error standard deviation, as we have less confidence on the correctness of their value. The Uncertain Moving Average (UMA) is based on the moving average, where sequence $S$ is substituted by $S_p$, and the point $pm_i$ is defined as follows:

$$pm_i = \frac{\sum_{j=i-w}^{i+w} \frac{v_j}{s_j}}{2w+1} \quad (17)$$

where $s_j$ is the standard deviation of random variable $t_j$.

The Uncertain Exponential Moving Average (UEMA) is based on the exponential moving average, where sequence $S$ is substituted by $S_e$, and point $pe_i$ is defined as follows:

$$pe_i = \frac{\sum_{j=i-w}^{i+w} v_j \frac{e^{-\lambda|j-i|}}{s_j}}{\sum_{j=i-w}^{i+w} e^{-\lambda|j-i|}} \quad (18)$$

At this point, we have introduced the UMA and UEMA filters. These filters allow us to reduce the signal noise, but do not define any distance function. In the subsequent experiments, we consider the Euclidean distance computed on the sequences filtered by UMA and UEMA techniques. Thus, Euclidean, UMA, and UEMA share the same distance function, but the input sequence is different.

### 5.2 Performance

We first examine the behavior of UMA and UEMA when we vary the parameters window size, $w$, and decaying factor, $\lambda$. Figure 13 depicts the effect of varying $w$ between $0-20$ on the $F_1$ score. The results are averaged over all datasets. Note that when $w=0$, UMA and UEMA degenerate to the simple Euclidean distance. We observe that the accuracy for UMA increases by 13% as we increase $w$ from 0 to 2, and then starts falling again as we further increase $w$. Evidently, aggregating many points (i.e., large $w$) is equally ineffective as not aggregating at all (i.e., $w=0$), since distant neighbors do not carry much (if at all) information about the current point.

The graphs also shows the performance of UEMA for two different $\lambda$ settings. For a small decaying factor, $\lambda = 0.1$,

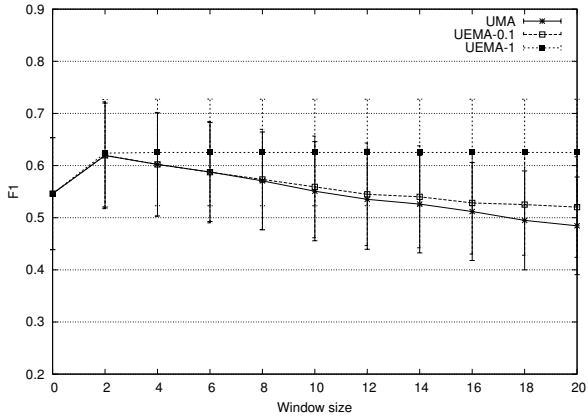

**Figure 13: F1 score varying the window size, $w$, for UMA and UEMA (with $\lambda = 0.1, 1$).**

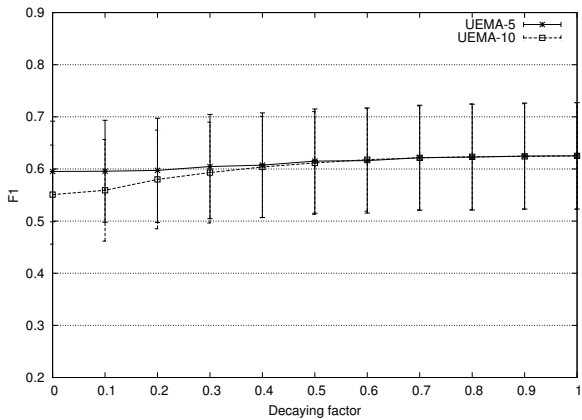

**Figure 14: F1 score varying the decaying factor, $\lambda$, for UEMA (for $w = 5, 10$).**

UEMA performs very close to UMA, since all the points in the window are assigned similar weights. This effect diminishes as $w$ increases and $\lambda$ introduces a higher variation among the weights of the near and distant neighbors of the current point. When we use a high value for the decaying factor, $\lambda = 1$, the effect of the distant neighbors diminishes much faster, thus, rendering the size of the window irrelevant for the performance of UEMA.

In Figure 14 we illustrate how the accuracy of UEMA varies when we change $\lambda$ (the case $\lambda = 0$ is equivalent to UMA). The experiments show that $\lambda$ has only a small effect on the performance of the algorithm, especially when the size of the window is small.

Overall, we note that UMA and UEMA exhibit a relatively stable behavior with respect to their parameters. For the rest of this study, we assume a decaying factor of $\lambda = 1$ for UEMA, and a moving average window length $W = 5$ (i.e., $w = 2$) for both UMA and UEMA.

In the next set of experiments, we compare the accuracy of Euclidean, DUST[4], UMA, and UEMA techniques on all

---

[4]Based on the previous experiments, DUST performs at least as good, or better than MUNICH and PROUD for

datasets perturbed with normal mixed error distribution, where 20% points with error standard deviation 1.0, and the remaining 80% with error standard deviation 0.4. This setting was chosen to stress-test the techniques. Every time series in each dataset was used as a query, and the results are averaged over all these time series.

Figure 15 depicts the results for the above experiment. The accuracy of DUST and Euclidean is almost the same, while UMA and UEMA perform consistently better, with the latter achieving the best performance among all techniques. Similar results were obtained for the uniform and exponential mixed error distributions, as shown in Figures 15 and 17, respectively.

The graphs show that (on average, across all datasets) Euclidean is always the worst performer, with a drop of 9% in its performance for the mixed exponential error distribution, which represents the hardest case. DUST performs close to Euclidean for the mixed normal and uniform distribution, but manages to maintain the same level of performance for the mixed exponential distribution as well.

UMA and UEMA exhibit the highest accuracy levels, averaging 4% to 15%, respectively, higher than DUST, and maintaining the same level of performance across all error distributions. Overall the $F_1$ score of UEMA is 4% higher than that of UMA.

The above results are very interesting: the intuitive and simple UMA and UEMA techniques outperform DUST, a complex method that requires much more a priori knowledge on the data distributions. Instead, these experiments indicate that much of the knowledge is conveyed in the error standard deviation, and in the distribution of the neighboring points. UMA and UEMA are the best performers, because they do *not* assume that data points are independent, a simplifying, yet unrealistic assumption made by the techniques previously proposed in the literature.

Note that UMA and UEMA are also computationally efficient, requiring almost the same time as Euclidean, and significantly less time than DUST, PROUD, and MUNICH. All the above observations indicate that UEMA is the method of choice for similarity matching in uncertain time series, when a probabilistic measure of certainty for the similarity is not required. Even when such a measure is required, UEMA can serve as a baseline for the target performance.

## 6. DISCUSSION

In this work, we reviewed the existing techniques for similarity matching in uncertain time series, and performed analytical and experimental comparisons of the techniques. Based on our evaluation, we can provide some guidelines for the use of these techniques.

MUNICH and PROUD are based on the Euclidean distance, while DUST proposes a new distance measure. Nevertheless, DUST outperforms Euclidean only if the distribution of the observation errors is mixed, and the parameters of this distribution are known.

An important factor for choosing among the available techniques is the information that is available about the distribution of the time series and its errors. When we do not have enough, or accurate information on the distribution of the error, PROUD and DUST do not offer an advantage

---

a variety of settings. Therefore, we only report the performance of DUST in these experiments for ease of exposition.

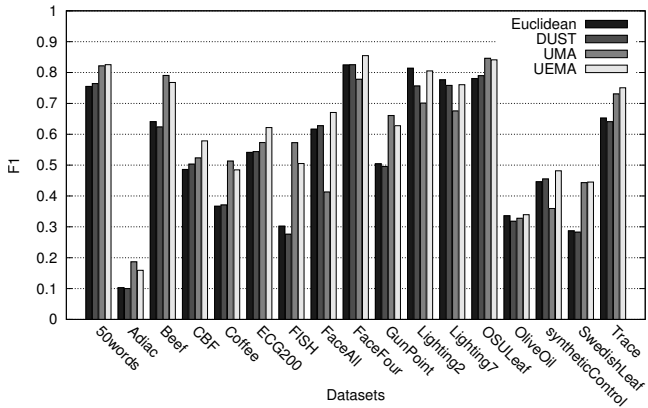

**Figure 15: F1 score for all datasets and mixed error distribution: uniform with 20% standard deviation 1.0, and 80% standard deviation 0.4.**

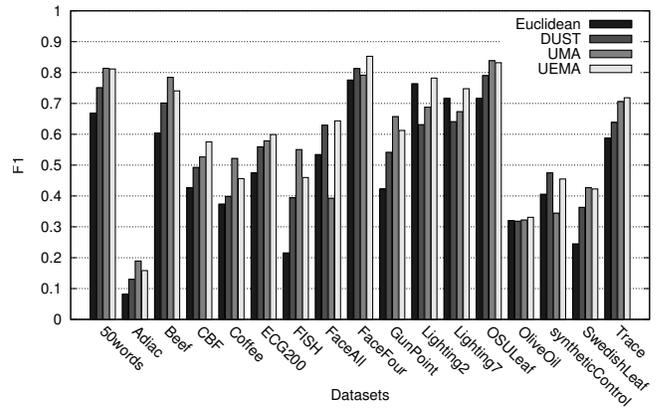

**Figure 17: F1 score for all datasets and mixed error distribution: exponential with 20% standard deviation 1.0, and 80% with standard deviation 0.4.**

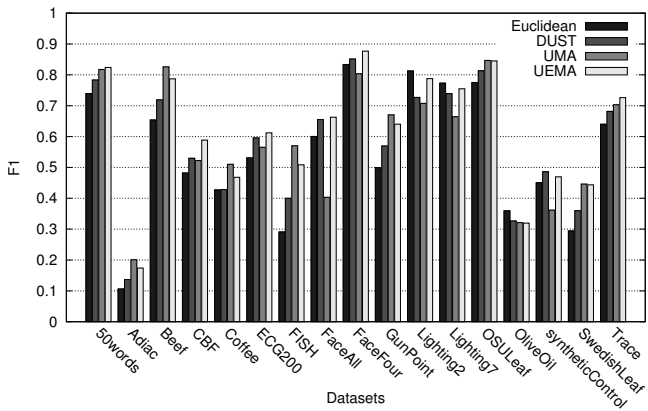

**Figure 16: F1 score for all datasets and mixed error distribution: normal with 20% standard deviation 1.0, and 80% with standard deviation 0.4.**

in terms of accuracy when compared to Euclidean. Nevertheless, Euclidean does not provide quality guarantees while MUNICH and PROUD do.

The probabilistic threshold $\tau$ has a considerable impact on the accuracy of the MUNICH and PROUD techniques. However, it not obvious how to set $\tau$, and no theoretical analysis has been provided on that. The only way to pick the correct value is by experimental evaluation, which can sometimes become cumbersome.

Our experiments showed that MUNICH is applicable only in the cases where the standard deviation of the error is relatively small, and the length of the time series is also small (otherwise the computational cost is prohibitive). However, we note that this may not be a restriction for some real applications. Indeed, MUNICH's high accuracy may be a strong point when deciding the technique to use.

The UMA and UEMA moving average filters proved to be very effective, outperforming the previous techniques in a variety of settings. This surprising result is due to the ability of the moving average to exploit the correlation of neighboring points in a very intuitive and simple manner: it reduces the effect of errors, which the filter levels out. Ignoring the strong correlation exhibited by neighboring points in the time series is not beneficial. Indeed, as our study shows, it is a severe limitation of all the techniques previously proposed in the literature.

However, we should note that the goal of MUNICH and PROUD is to provide an additional probabilistic measure of certainty for the computed similarity value. This is something that we cannot readily get from UMA, UEMA, or DUST, and may be important for certain applications.

Finally, we observe that there exist some datasets for which all techniques perform well (e.g., FaceFour and OSU-Leaf), and others for which accuracy is low (e.g., Adiac and Swedish Leaf). A close look at the characteristics of these datasets revealed that datasets for which the average distance between time series was low led to low accuracy. This is because uncertainty has a significant impact for these datasets, making it hard to distinguish the time series and select a clear winner for the similarity matching problem. On the other hand, the same level of uncertainty does not affect much datasets that have a high average distance among their time series.

## 7. CONCLUSIONS

The emerging area of uncertain time series processing and analysis is increasingly attracting the attention of both the research community and the practitioners in the field, mainly because of the applications and interesting problems it entails.

In this study, we evaluated the state of the art techniques for similarity matching in uncertain time series, as this operation is the basis for more complex algorithms. Apart from the techniques that were previously proposed in the literature, we also evaluated two additional, obvious alternatives that were not studied before.

Our experiments were based on 17 real, diverse datasets, and the results demonstrate that simple measures, based on moving average, outperform the more sophisticated alternatives. These results also suggest that a promising direction is to develop measures that take into account the sequential correlations inherent in time series.

## Acknowledgments

We would like to thank the reviewers for their useful comments and suggestions. Part of this work was supported by the FP7 EU IP project KAP (grant agreement no. 260111).

## 8. REFERENCES

[1] Keogh, E., Xi, X., Wei, L. & Ratanamahatana, C. A. (2006). The UCR Time Series Classification/Clustering Homepage: www.cs.ucr.edu/~eamonn/time_series_data/. Accessed on 17 May 2011.

[2] C. Aggarwal. On Unifying Privacy and Uncertain Data Models. In *ICDE*. IEEE, 2008.

[3] C. Aggarwal. *Managing and Mining Uncertain Data*. Springer-Verlag New York Inc, 2009.

[4] R. Agrawal, C. Faloutsos, and A. Swami. Efficient similarity search in sequence databases. *Foundations of Data Organization and Algorithms*, 1993.

[5] J. Aßfalg, H.-P. Kriegel, P. Kröger, and M. Renz. Probabilistic similarity search for uncertain time series. In *SSDBM*, pages 435–443, 2009.

[6] D. J. Berndt and J. Clifford. Using dynamic time warping to find patterns in time series. In *KDD Workshop*, pages 359–370, 1994.

[7] M. Ceriotti, M. Corra, L. D'Orazio, R. Doriguzzi, D. Facchin, S. Guna, G. P. Jesi, R. L. Cigno, L. Mottola, A. L. Murphy, M. Pescalli, G. P. Picco, D. Pregnolato, and C. Torghele. Is There Light at the Ends of the Tunnel? Wireless Sensor Networks for Adaptive Lighting in Road Tunnels. In *International Conference on Information Processing in Sensor Networks (IPSN)*, pages 187–198, 2011.

[8] K. Chan and A. Fu. Efficient time series matching by wavelets. In *ICDE*, pages 126–133. IEEE, 2002.

[9] H. Ding, G. Trajcevski, P. Scheuermann, X. Wang, and E. Keogh. Querying and mining of time series data: experimental comparison of representations and distance measures. *PVLDB*, 2008.

[10] C. Faloutsos, M. Ranganathan, and Y. Manolopoulos. Fast subsequence matching in time-series databases. *SIGMOD Conference*, 23(2):419–429, 1994.

[11] B. C. M. Fung, K. Wang, R. Chen, and P. S. Yu. Privacy-preserving data publishing: A survey of recent developments. *ACM Comput. Surv.*, 42(4), 2010.

[12] M. Hamilton, E. Graham, P. Rundel, M. Allen, W. Kaiser, M. Hansen, and D. Estrin. New Approaches in Embedded Networked Sensing for Terrestrial Ecological Observatories. *Environmental Engineering Science*, 24(2), 2007.

[13] E. Keogh, K. Chakrabarti, M. Pazzani, and S. Mehrotra. Dimensionality reduction for fast similarity search in large time series databases. *Knowledge and Information Systems*, 3(3):263–286, 2001.

[14] L. Krishnamurthy, R. Adler, P. Buonadonna, J. Chhabra, M. Flanigan, N. Kushalnagar, L. Nachman, and M. Yarvis. Design and deployment of industrial sensor networks: experiences from a semiconductor plant and the north sea. In *Embedded networked sensor systems*, pages 64–75. ACM, 2005.

[15] X. Lian and L. Chen. Efficient join processing on uncertain data streams. In *CIKM*, pages 857–866, 2009.

[16] J. Lin, E. Keogh, L. Wei, and S. Lonardi. Experiencing SAX: a novel symbolic representation of time series. *Data Mining and Knowledge Discovery*, 15(2):107–144, 2007.

[17] C. Ma, H. Lu, L. Shou, G. Chen, and S. Chen. Top-$k$ similarity search on uncertain trajectories. In *SSDBM*, pages 589–591, 2011.

[18] Y. Moon, K. Whang, and W. Han. General match: a subsequence matching method in time-series databases based on generalized windows. In *SIGMOD Conference*, pages 382–393. ACM, 2002.

[19] Y. Moon, K. Whang, and W. Loh. Duality-based subsequence matching in time-series databases. In *ICDE*, pages 263–272. IEEE, 2002.

[20] S. Papadimitriou, F. Li, G. Kollios, and P. S. Yu. Time series compressibility and privacy. In *VLDB*, 2007.

[21] U. Raza, A. Camerra, A. L. Murphy, T. Palpanas, and G. P. Picco. What does model-driven data acquisition really achieve in wireless sensor networks? In *IEEE International Conference on Pervasive Computing and Communications*, Lugano, Switzerland, 2012.

[22] N. Roussopoulos, S. Kelley, and F. Vincent. Nearest neighbor queries. In *SIGMOD Conference*, pages 71–79. ACM, 1995.

[23] S. Sarangi and K. Murthy. DUST: a generalized notion of similarity between uncertain time series. In *SIGKDD*, pages 383–392. ACM, 2010.

[24] J. Shieh and E. J. Keogh. $i$SAX: indexing and mining terabyte sized time series. In *KDD*, pages 623–631, 2008.

[25] M. Stonebraker, J. Becla, D. J. DeWitt, K.-T. Lim, D. Maier, O. Ratzesberger, and S. B. Zdonik. Requirements for science data bases and scidb. In *CIDR*, 2009.

[26] D. Suciu, A. Connolly, and B. Howe. Embracing uncertainty in large-scale computational astrophysics. In *MUD*, pages 63–77, 2009.

[27] G. Trajcevski, A. N. Choudhary, O. Wolfson, L. Ye, and G. Li. Uncertain range queries for necklaces. In *Mobile Data Management*, pages 199–208, 2010.

[28] T. T. L. Tran, L. Peng, B. Li, Y. Diao, and A. Liu. Pods: a new model and processing algorithms for uncertain data streams. In *SIGMOD Conference*, pages 159–170, 2010.

[29] M. Yeh, K. Wu, P. Yu, and M. Chen. PROUD: a probabilistic approach to processing similarity queries over uncertain data streams. In *EDBT*, pages 684–695. ACM, 2009.

[30] Y. Zhao, C. C. Aggarwal, and P. S. Yu. On wavelet decomposition of uncertain time series data sets. In *CIKM*, pages 129–138, 2010.